# Asperity Shape and Gradient Elasticity in Flexoelectric/Triboelectric Contacts

Karl P. Olson[a], Laurence D. Marks[a*]

[a]Northwestern University, Department of Materials Science and Engineering
2220 Campus Drive, Room 2036
Evanston, IL, United States 60208

*Corresponding Author, l-marks@northwestern.edu

**Abstract**

The underlying mechanisms responsible for triboelectricity have yet to be completely understood. We have previously proposed a model which explains charge transfer in non-metals via band bending due to electromechanical, especially flexoelectric, effects at deformed asperities coupled with work function differences. Here, we investigate whether the shape of asperities is important for triboelectricity. The results indicate that the shape is important in general, since how the electromechanical response scales with force and asperity size depends on the shape. This is qualitatively in agreement with experimental results. Further, we discuss how the impact of the shape depends on material, geometric, and electronic transport details. Additionally, gradient elasticity is incorporated into the model. As asperity contact is a nanoscale phenomenon, size-dependent mechanics can become significant and give more physically reasonable results. In some cases, the impact of gradient elasticity terms on the electromechanical potentials is very large, indicating that standard elasticity theory is not enough to cover some relevant cases in modelling triboelectricity.

# 1 Introduction

Triboelectricity, the charge transfer between two contacting or rubbing materials, is of interest in a wide range of scientific, engineering, and everyday applications. In some, such as in the design of triboelectric nanogenerators (TENGs) (Fan et al., 2012; Kim et al., 2021), large charge transfer is desirable. In others, less charge transfer is preferable, e.g., to prevent industrial accidents (Liang et al., 1996), pharmaceutical powder clumping (Naik et al., 2016; Watanabe et al., 2007; Wong et al., 2015), or excessive charging of space exploration vehicles (Mishra and Sana, 2022). Triboelectricity is also important in diverse scientific areas, from planetary formation (Steinpilz et al., 2019) to the effect of shampoos on the static electricity of human hair (Mills et al., 1956).

The need for fundamental knowledge of triboelectricity is clear. Despite an increasing interest in triboelectric research, many details of the mechanism and relevant materials and geometrical properties are still unknown (Lacks and Shinbrot, 2019; Pan and Zhang, 2019; Williams, 2012). Previously, we have shown that triboelectricity occurs at contacting asperities, where electromechanical potentials large enough to drive charge transfer form due to asperity deformation (Mizzi et al., 2019; Mizzi and Marks, 2022). A significant contribution is due to the flexoelectric effect, the polarization caused by strain gradients.

Going beyond this, we have also developed a more detailed model to analyze the electromechanics of single-asperity contacts (Olson et al., 2022). The contact of asperities is simplified to a spherical metal indenter that is pressed into a semi-infinite semiconductor slab. The deformation is determined using Hertzian contact solutions (Hertz, 1882), and the electronic band bending is calculated by considering electromechanical and other relevant electronic effects.

While this model explained changes in electronic transport for a platinum-iridium ($Pt_{0.8}Ir_{0.2}$) - Nb-doped strontium titanate ($Nb:SrTiO_3$) system, it included assumptions that are not valid in general. One point concerns the asperity shape. In tribology, the shape of asperities is in some cases considered to have little importance, such as its effect on real contact areas (Bhushan, 1998; Greenwood et al., 1966) or on friction (Siripuram and Stephens, 2004). In other cases, tribological quantities of interest, such as leakage from a hydrodynamic film (Siripuram and Stephens, 2004), creep at asperities (Alamos et al., 2021), or wear rate in abrasives (De Pellegrin and Stachowiak, 2004), are dependent on the asperity shape.

With a focus on triboelectricity, the effect of asperity shape has been studied mainly in the context of surface modifications of TENGs (Aazem et al., 2022; Zou et al., 2021). That is, microscale patterns are created on the surface of materials for example by molding polymers to solid templates (Varghese et al., 2022; Zhang et al., 2018) or selectively melting polymer surfaces with lasers (Huang et al., 2019; Muthu et al., 2020). Results in this area indicate that these surface modifications such as grating or arrays of protrusions or divots often increase charge transfer, thereby enhancing TENG performance. One work compared surfaces with regular arrays of microscale domes and pyramids to flat surfaces (Varghese et al., 2022). TENG open-circuit voltage and short-circuit current were increased by ~50% for domes and ~600% for pyramids. These results are attributed to increases in contact area, but this does not seem to entirely explain the massive increase in charge transfer, as with complete contact, the domes and pyramids would increase the contact area by ~60% and ~20%, respectively. The difference in the electromechanics resulting from the different asperity shapes may in fact be a significant part of the contribution. Various other works have examined the effect of different surface morphologies on the performance of TENGs (Mahmud et al., 2016; Tcho et al., 2017; Zhang et al., 2018). Consistently, the performance difference is contributed to contact area changes and sometimes changes in friction, even though charge transfer increases are often much larger than the changes in contact area. We will return to this at the end of this manuscript.

In the present work, we consider the significance of the asperity shape and size for triboelectricity from the perspective of electromechanical band bending generated at contacts. The results clearly show that the electromechanical effects of, e.g., sphere and cone indenter contacts and contacts of different sizes are different beyond the simple difference in contact areas. These results explain some of the results of experiments with surface-modified TENGs, for which surface area changes alone are not adequate.

Another issue extending beyond the previous model concerns the size-dependence of the system. Since asperities are on the nanometer scale, strain gradient elasticity may become important in some cases. Strain gradient elasticity is a higher-order elasticity theory that includes a strain gradient term in the strain energy expression, which is required to accurately describe size effects of deformation on the micro- or nanoscale (Lam et al., 2003). Here, we explore the effects of strain gradient elasticity on the triboelectric contact problem and describe cases for which it must be considered.

## 2 Theory

In the present work, we examine the effect of the asperity shape and strain gradient elasticity on the electromechanical potentials, i.e., the effects on the electronic band bending of strain, via the deformation potential (Stengel, 2015) and the mean-inner potential (Mizzi and Marks, 2021), and of strain gradients, via flexoelectricity. Depending on the materials involved in specific problems, other potentials may be necessary to include. For example, the Nb:SrTiO$_3$ – Pt$_{0.8}$Ir$_{0.2}$ case (Olson et al., 2022) requires consideration of the depletion potential. For this work, these material system dependent potentials are left aside. Unless otherwise noted, the calculations presented here consider a rigid indenter contacting a SrTiO$_3$ half space. That is, quantities such as the deformation potential, mean inner potential, Poisson's ratio, and the relative values of the elastic tensor components are assumed to have the values of SrTiO$_3$.

To investigate the effects of asperity shape, we consider five cases as shown in Fig. 1. The 3-D axisymmetric cases are sphere, cylinder, and cone indenters, as shown in Fig. 1(a)-(c). The 2-D cases with axial symmetry correspond to indenters in the shape of an infinitely long cylinder or rectangular prism, as shown in Fig. 1(d)-(e). These five indenters will be referred to as the sphere, cylinder, cone, roller, and punch, respectively. While real asperity shapes are intermediate to these, they provide bounds to what can be expected for more realistic shapes (De Pellegrin and Stachowiak, 2004).

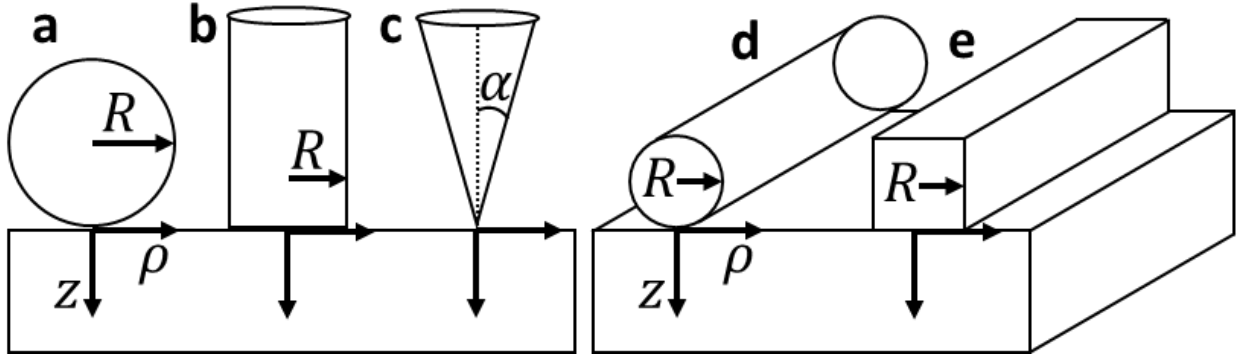

Fig. 1. Sketches, axis definitions, and geometric parameters for the five indenter cases. (a) sphere, (b) cylinder, (c) cone, (d) roller, and (e) punch.

Previously, we have developed a model to calculated the band bending of the conduction band of a Nb:SrTiO$_3$ half-space indented by a Pt$_{0.8}$Ir$_{0.2}$ sphere (Olson et al., 2022). From this work, the position of the conduction band of SrTiO$_3$, $E_c$, with respect to the Fermi level $E_F$, is

$$E_c(F, \mathbf{r}) - E_F = (D_{BS}^C + \varphi)\varepsilon_{\text{vol}}(F, \mathbf{r}) + \Phi_{\text{FXE}}(F, \mathbf{r}) + \Phi_{\text{DEP}}(\mathbf{r}) \tag{1a}$$

$$\Phi_{\text{FXE}}(F, \mathbf{r}) = \frac{q}{4\pi} \int_\Omega f_{ijkl} \frac{d\varepsilon_{kl}(F, \mathbf{r}')}{dr_j'} \cdot \frac{\mathbf{r} - \mathbf{r}'}{|\mathbf{r} - \mathbf{r}'|^3} d\mathbf{r}' \tag{1b}$$

where $\mathbf{r} = (\rho, \theta, z)$ is a 3-D cylindrical coordinate system or $\mathbf{r} = (\rho, z)$ a 2-D Cartesian coordinate system, $F$ the vertical force on the indenter in the 3-D cases or the force per unit length in the 2-D cases, $D_{BS}^C$ and $\varphi$ the conduction-band-specific deformation potential described by Stengel (Stengel, 2015) and the change in the mean inner potential due to strain (Mizzi and Marks, 2021), respectively, $\Phi_{\text{DEP}}$ the depletion potential (Yamamoto et al., 1998), $\varepsilon_{\text{vol}}$ the volumetric strain, $\Phi_{\text{FXE}}$ the flexoelectric potential, $q$ the elementary charge, $f_{ijkl}$ the flexocoupling voltages (i.e., the flexoelectric coefficients normalized by the dielectric constant (Hong and Vanderbilt, 2013)), $\varepsilon_{kl}$ the symmetrized strain tensor, and $\Omega$ the sample volume. As we noted above, we will proceed ignoring the $\Phi_{\text{DEP}}$ term.

The stresses throughout the half-space for each indenter shape are calculated from the Hertzian solutions for, respectively, a sphere of radius $R$, a cylinder of radius $R$, a cone with half-angle $\alpha$ measured from the vertical, a roller of radius $R$, and a punch of half-width $R$, all rigid and indenting an elastic half-space (Fischer-Cripps, 2007; Johnson, 1985; M'Ewen, 1949). The Hertzian solutions assume a pressure distribution at the surface limited to the region inside the contact radius, given in equation 2, where $p_m$ is the mean pressure at the surface, $a$ the contact radius, and $Y$ the Young's modulus of the elastic body. Note that none of these pressure distributions are differentiable at the edge of the contact region, $\rho = a$ (i.e., the pressure distributions are not smooth at $\rho = a$). While this is physically unreasonable, Hertzian solutions nonetheless give results with acceptable error for many contact problems (Johnson, 1982).

$$\sigma_z^{\text{sphere}}(z = 0) = -\frac{3}{2} p_m \left(1 - \frac{\rho^2}{a^2}\right)^{\frac{1}{2}}; \quad \rho \leq a = \left(\frac{3}{4} \frac{RF(1-\nu^2)}{Y}\right)^{\frac{1}{3}} \tag{2a}$$

$$\sigma_z^{\text{cylinder}}(z=0) = -\frac{1}{2}p_m\left(1-\frac{\rho^2}{a^2}\right)^{-\frac{1}{2}}; \quad \rho \leq a = R \tag{2b}$$

$$\sigma_z^{\text{cone}}(z=0) = -p_m \cosh^{-1}\frac{a}{\rho}; \quad \rho \leq a = \left(\frac{2(1-\nu^2)F\tan\alpha}{Y}\right)^{\frac{1}{2}} \tag{2c}$$

$$\sigma_z^{\text{roller}}(z=0) = -\frac{4}{\pi}p_m\left(1-\frac{x^2}{a^2}\right)^{\frac{1}{2}}; \quad \rho \leq a = \left(\frac{4(1-\nu^2)FR}{\pi E}\right)^{\frac{1}{2}} \tag{2d}$$

$$\sigma_z^{\text{punch}}(z=0) = -\frac{2}{\pi}p_m\left(1-\frac{x^2}{a^2}\right)^{-\frac{1}{2}}; \quad \rho \leq a = R \tag{2e}$$

The results of these calculations are, of course, dependent on $F$, $R$ or $\alpha$, and $Y$, but some generalizations and reductions are possible.

1. The mean contact pressure is given by $p_m = F/\pi a^2$ for the 3-D cases and by $p_m = F/2a$ for the 2-D cases.
2. The stress inside the half-space is proportional to $p_m$, so the strain is proportional to $p_m/Y$.
3. The flexoelectric potential $\Phi_{\text{FXE}}$ scales as the strain.
4. Normalizing the coordinates **r** by the contact radius $a$ results in a natural scaling.

Specifically, $\Phi_{\text{FXE}}$ scales as:

$$\Phi_{\text{FXE}}^{\text{sphere}}\left(F,R,Y,\nu,\boldsymbol{\mu},\frac{\mathbf{r}}{a}\right) \propto \frac{p_m^{\text{sphere}}}{Y} = \left(\frac{16F}{9\pi^3(1-\nu^2)^2 R^2 Y}\right)^{\frac{1}{3}} \propto \left(\frac{F}{R^2 Y}\right)^{\frac{1}{3}} \tag{3a}$$

$$\Phi_{\text{FXE}}^{\text{cylinder}}\left(F,R,Y,\nu,\boldsymbol{\mu},\frac{\mathbf{r}}{a}\right) \propto \frac{p_m^{\text{cylinder}}}{Y} = \frac{F}{\pi R^2 Y} \propto \frac{F}{R^2 Y} \tag{3b}$$

$$\Phi_{\text{FXE}}^{\text{cone}}\left(F,R,Y,\nu,\boldsymbol{\mu},\frac{\mathbf{r}}{a}\right) \propto \frac{p_m^{\text{cone}}}{Y} = \frac{1}{2(1-\nu^2)\tan\alpha} \propto \frac{1}{\tan\alpha} \tag{3c}$$

$$\Phi_{\text{FXE}}^{\text{roller}}\left(F,R,Y,\nu,\boldsymbol{\mu},\frac{\mathbf{r}}{a}\right) \propto \frac{p_m^{\text{roller}}}{Y} = \left(\frac{\pi F}{16RY(1-\nu^2)}\right)^{\frac{1}{2}} \propto \left(\frac{F}{RY}\right)^{\frac{1}{2}} \tag{3d}$$

$$\Phi_{\text{FXE}}^{\text{punch}}\left(F,R,Y,\nu,\boldsymbol{\mu},\frac{\mathbf{r}}{a}\right) \propto \frac{p_m^{\text{punch}}}{Y} = \frac{F}{2RY} \propto \frac{F}{RY} \tag{3e}$$

where $\nu$ is the Poisson's ratio and $\boldsymbol{\mu}$ is the flexoelectric coefficient tensor.

Equation 3 includes only the parameters that lead to simple scaling. Because each stress component depends differently on $\nu$ (Fischer-Cripps, 2007; Johnson, 1985; M'Ewen, 1949), so do the strain components depend on $\nu$ in a complex manner. Additionally, how much each strain component impacts the result depends on the ratios of different flexoelectric tensor components. Therefore, $\Phi_{\text{FXE}}$ does not scale simply with $\nu$ or with changes in individual flexoelectric tensor components, beyond (see equation 1(b)) that the contribution is linear in the components. The effect of separate flexoelectric tensor components is discussed later, in section 3.3.

Because $\Phi_{\text{FXE}}$ scales as the strain, and the mean inner potential and band-specific deformation potential terms scale in the same way, we can report a normalized electromechanical potential

$\Phi_{EM}$ that includes the flexoelectric terms in addition to the other strain-dependent terms, given by equation 4,

$$\Phi_{EM}(v, \boldsymbol{\mu}, \frac{\mathbf{r}}{a}) = \left[\Phi_{FXE}\left(F, R, Y, v, \boldsymbol{\mu}, \frac{\mathbf{r}}{a}\right) + (\varphi + D_{BS}^{C})\varepsilon_{vol}\left(F, R, Y, v, \boldsymbol{\mu}, \frac{\mathbf{r}}{a}\right)\right] G(F, R, Y) \quad (4)$$

where $G(F, R, Y)$ is dependent on the shape and is the inverse of the rightmost term of equation 3. For example, $G(F, R, Y) = (R^2 Y/F)^{1/3}$ for sphere indenters. While $\Phi_{EM}$ is appropriately normalized, it is not unitless. The units are shape dependent and can be inferred from equations 3 and 4. To maintain consistent scaling, it necessarily uses normalized, unitless axes $\rho/a$ and $z/a$. As we have mentioned before, we will report $\Phi_{EM}$ using the values of SrTiO$_3$ for $v$ and $\boldsymbol{\mu}$.

## 3 Results

### 3.1 Electromechanical Potentials and Normalized Voltages

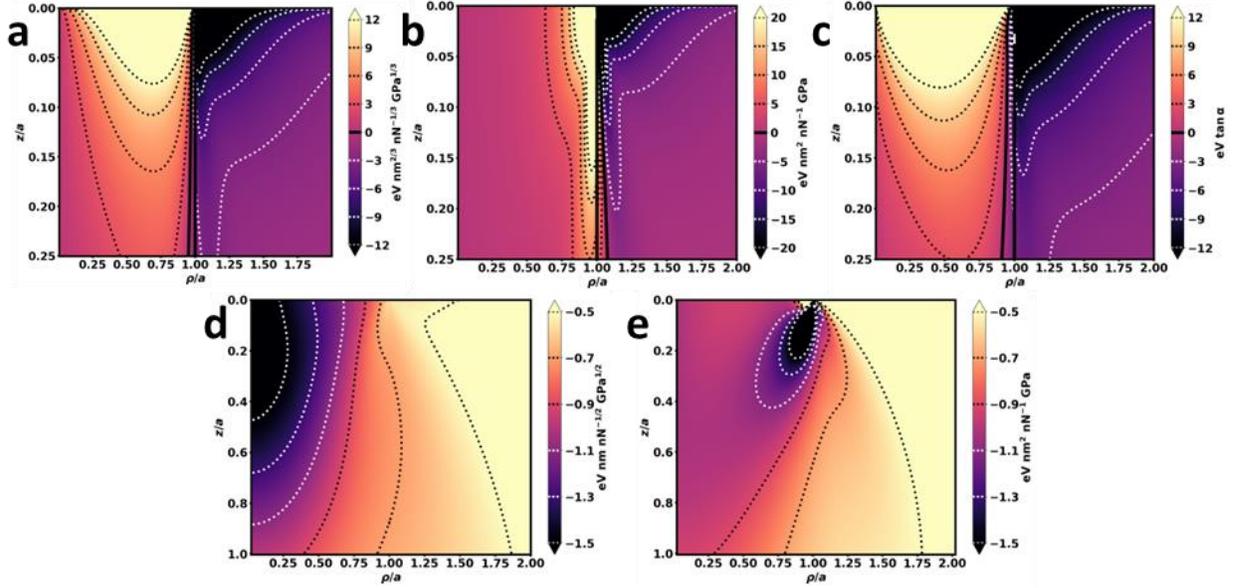

Fig. 2. Plots of $\Phi_{EM}$ for different indenter shapes: (a) sphere, (b) cylinder (c) cone, (d) roller, and (e) punch. $\Phi_{EM}$ is normalized as described in equation 4 and the $\rho$ and $z$ axes are normalized by the contact radius $a$. Contour lines are labeled in the color bar.

Fig. 2 shows $\Phi_{EM}$ plotted against normalized coordinates $\mathbf{r}/a$ for the five indenter shapes. In all the axisymmetric cases, the potential is positive inside the contact edge and negative outside – for the flexoelectric coefficients of SrTiO$_3$. Additionally, the potential outside the contact edge ($\rho > a$) has a roughly similar shape, while the shape of the positive region inside ($\rho < a$) is highly dependent on the shape of the indenter. On the other hand, the 2-D cases are distinct from the axisymmetric cases and from each other across all regions.

Understanding two quantities involved in the electromechanical contact can give further context and insight into the problem. We define the strain energy density $E'_{strain}(\mathbf{r}) = \sigma_{ij}(\mathbf{r})\varepsilon_{ij}(\mathbf{r})/2$. Like $\Phi_{FXE}$, the stress and the strain are proportional to $p_m$ as given by equation 3 when plotted against the normalized axes $\mathbf{r}/a$. Therefore, the total strain energy integrated over the whole slab volume $\Omega$ is $E_{\text{strain}} = \int_\Omega E'_{\text{strain}} \, d\mathbf{r}$, which is proportional to $\frac{p_m^2 a^3}{Y}$. For example, the total strain energy for the sphere is $E_{\text{strain}}^{\text{sphere}} \propto p_m^{\text{sphere}} \frac{p_m^{\text{sphere}}}{Y} a_{\text{sphere}}^3 \propto \left(\frac{FY^2}{R^2}\right)^{\frac{1}{3}} \left(\frac{F}{R^2Y}\right)^{\frac{1}{3}} \frac{RF}{Y} = \left(\frac{F^5}{RY^2}\right)^{\frac{1}{3}}$.

We can also define a normalized flexoelectric potential $V_{\text{FXE}} = \frac{1}{qa^3} \int_\Omega \Phi_{\text{FXE}} \, d\mathbf{r}$ where $\Omega$ is the slab volume (taken to be $\rho < 2a$ and $z < a$) and we note that the volume over which $\Phi_{\text{EM}}$ is not approximately zero is proportional to $a^3$. We can further define $V_{\text{FXE}}^{\text{inside}} = \frac{1}{qa^3} \int_{\Omega^{\text{inside}}} \Phi_{\text{EM}} \, d\mathbf{r}$ and $V_{\text{EM}}^{\text{outside}} = \frac{1}{qa^3} \int_{\Omega^{\text{outside}}} \Phi_{\text{FXE}} \, d\mathbf{r}$, where $\Omega^{\text{inside}}$ is the slab volume underneath the contact area ($\rho < a, z < a$) and $\Omega^{\text{outside}}$ is the slab volume outside the contact area, here considered to be ($a < \rho < 2a, z < a$).

$V_{\text{FXE}}$ should be expected to scale like $\Phi_{\text{FXE}}$ (see equation 3). Indeed, this is true – for example, $V_{\text{FXE}}^{\text{sphere}} \propto \left(\frac{F}{R^2Y}\right)^{\frac{1}{3}}$. The results for the various indenter shapes are given in equation 5.

$$V_{\text{FXE}}^{\text{sphere}} \left(\frac{F}{R^2Y}\right)^{-\frac{1}{3}} = -5.5 \text{ V}; \; V_{\text{FXE}}^{\text{sphere,inside}} \left(\frac{F}{R^2Y}\right)^{-\frac{1}{3}} = 12.1 \text{ V}; \; V_{\text{FXE}}^{\text{sphere,outside}} \left(\frac{F}{R^2Y}\right)^{-\frac{1}{3}} = -17.6 \text{ V} \quad (5a)$$

$$V_{\text{FXE}}^{\text{cylinder}} \left(\frac{F}{R^2Y}\right)^{-1} = 2.3 \text{ V}; \; V_{\text{FXE}}^{\text{cylinder,inside}} \left(\frac{F}{R^2Y}\right)^{-1} = 17.9 \text{ V}; \; V_{\text{FXE}}^{\text{cylinder,outside}} \left(\frac{F}{R^2Y}\right)^{-1} = -15.6 \text{ V} \quad (5b)$$

$$V_{\text{FXE}}^{\text{cone}} \tan \alpha = -13.4 \text{ V}; \; V_{\text{FXE}}^{\text{cone,inside}} \tan \alpha = 10.8 \text{ V}; \; V_{\text{FXE}}^{\text{cone,outside}} \tan \alpha = -24.1 \text{ V} \quad (5c)$$

Because the largest contributions to $V_{\text{FXE}}$ come from the abrupt gradients at the contact edge, the values in equation 5 are somewhat sensitive to the resolution of the discrete calculations. $V_{\text{FXE}}^{\text{inside}}$ and $V_{\text{FXE}}^{\text{outside}}$ may vary by a few percent if care is not taken, and because those are large but have opposite signs, $V_{\text{FXE}}$ may vary as much as 10 or 20 percent. Calculations of other quantities presented here, such as the plots of $\Phi_{\text{EM}}$, are much better behaved and do not strongly depend on calculation parameters, as long as the stresses, potentials, etc. are very small at the $\rho$ and $z$ boundaries, far away from the contact.

### Section 3.2 Elastic Spherical Indenters

Above, we have considered rigid indenters. For an elastic sphere indenter, the results are quite similar. In this case, equation 3(a) is modified to

$$\Phi_{\text{FXE}}^{\text{sphere}}(F, R, Y_*, Y, \mathbf{r}/a) \propto p_m^{\text{sphere}}/Y = \left(\frac{16FY_*^2}{9\pi^3 R^2 Y^3}\right)^{\frac{1}{3}} \propto \left(\frac{FY_*^2}{R^2 Y^3}\right)^{\frac{1}{3}} \quad (6)$$

and the right side of equation 2(a) to $a = \left(\frac{3}{4}\frac{RF}{Y_*}\right)^{\frac{1}{3}}$, where $Y_* = \left(\frac{1-\nu^2}{Y} + \frac{1-\nu_i^2}{Y_i}\right)^{-1}$ is the reduced modulus and the half-space and indenter have, respectively, Young's moduli $Y$ and $Y_i$ and Poisson's ratios $\nu$ and $\nu_i$. The contact pressure, and therefore stress, within the half-space is dependent on the reduced modulus, while the strain due to this stress is dependent only on the modulus of the half-space. To convert from the $\Phi_{\text{EM}}$ calculated for a rigid sphere indenter to the result for an elastic sphere indenter, a simple multiplication by $\left(\frac{Y_*(1-\nu^2)}{Y}\right)^{\frac{2}{3}}$ is sufficient. The same factor is also appropriate for the flexoelectric potential $V_{\text{FXE}}$, while the strain energy $E_{\text{strain}}$ scaling should be multiplied by the factor $\left(\frac{Y_*(1-\nu^2)}{Y}\right)^{\frac{1}{3}}$. Elastic indenter results for other shapes are not so simple since the unphysically sharp corners of other indenters directly interact with the half-space (Popov et al., 2019).

### 3.3 Contribution of Flexoelectric Tensor Components

A more generalized analysis can be realized by considering $\Phi_{\text{FXE}}$ and $(D_{BS}^C + \varphi)\varepsilon_{\text{vol}}(F, \mathbf{r})$ separately rather than their combination, $\Phi_{\text{EM}}$. Further, we can consider three separate components of $\Phi_{\text{FXE}}$ that each depend on only one of the three independent flexoelectric coefficients in a cubic system. That is, we artificially set all but one flexoelectric coefficient to zero and calculate $\Phi_{\text{FXE}}$. For a rigid sphere indenter, the result is plotted in Fig. 3, which has three cases: non-zero longitudinal $\mu_{1111} = \mu'$, non-zero transverse $\mu_{1122} = \mu'$, and non-zero shear $\mu_{1212} = \mu'$ coefficients. Here, $\Phi_{\text{FXE}}$ is normalized by terms that scale with strain as described for $\Phi_{\text{EM}}$ in equation 4 and is further normalized by the non-zero flexoelectric coefficient $\mu'$. We notate this normalized quantity as $\Phi_{\text{FXE}}^{\text{NORM}}$. $\varepsilon_{\text{vol}}$ is also normalized by the terms described in equation 4. $\Phi_{\text{EM}}$ can be obtained by simply adding in the appropriate proportions the $\Phi_{\text{FXE}}^{\text{NORM}}$ terms and $(D_{BS}^C + \varphi)\varepsilon_{\text{vol}}(F, \mathbf{r})$, as shown in equation 7.

$$\Phi_{\text{EM}} = \mu_{1111}\Phi_{FXE}^{\text{longitudinal}} + \mu_{1122}\Phi_{FXE}^{\text{transverse}} + \mu_{1212}\Phi_{FXE}^{\text{shear}} + (D_{BS}^C + \varphi)\varepsilon_{\text{vol}} \qquad (7)$$

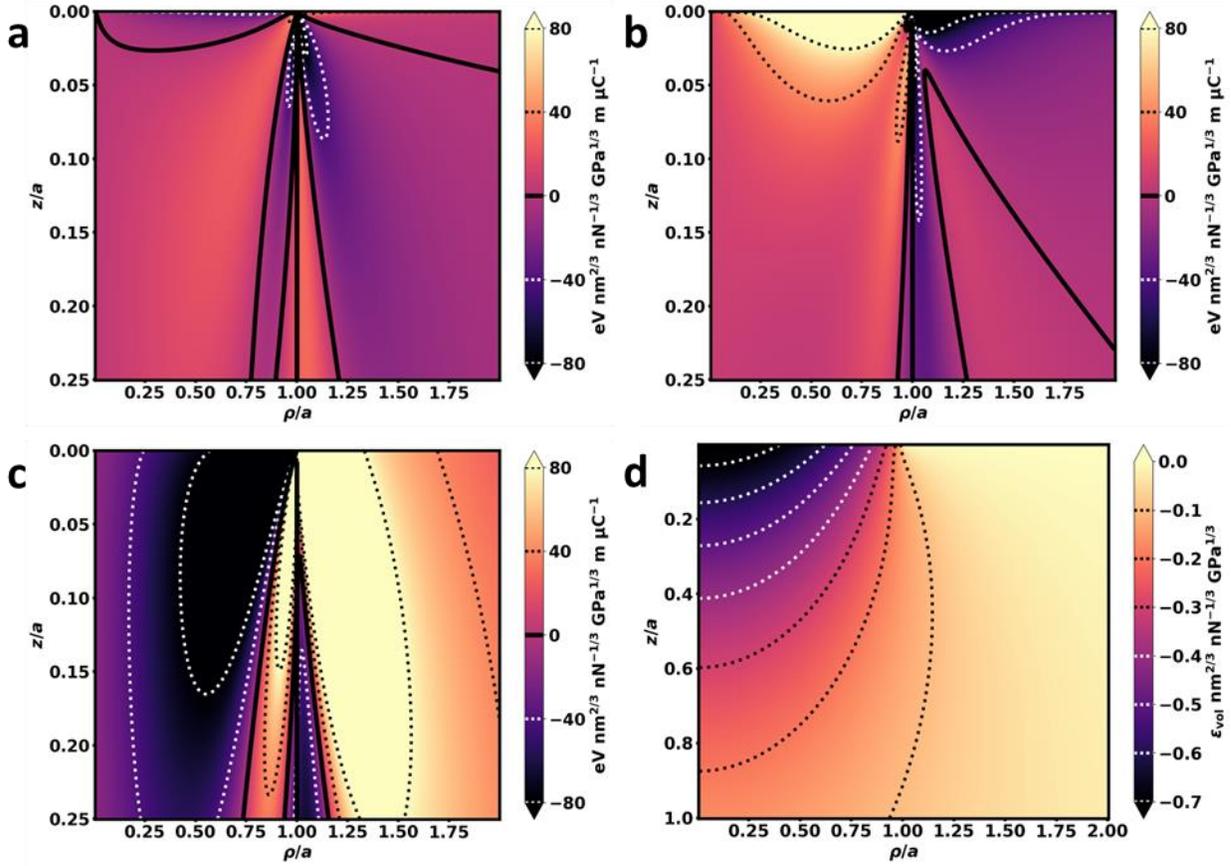

Fig. 3. (a-c) Normalized $\Phi_{\text{FXE}}^{\text{NORM}}$ for a rigid sphere indenter with artificial flexoelectric coefficients, where $\mu'$ is a constant flexoelectric coefficient. (a) $\mu_{1111} = \mu'$, $\mu_{1122} = \mu_{1212} = 0$. (b) $\mu_{1122} = \mu'$, $\mu_{1111} = \mu_{1212} = 0$. (c) $\mu_{1212} = \mu'$, $\mu_{1111} = \mu_{1122} = 0$. (d) Normalized $\varepsilon_{\text{vol}}$ for a rigid sphere indenter.

Similarly, the total flexoelectric potential given in equation 5 can be split into its components and normalized by the flexoelectric coefficients. The three components contribute $V_{\text{FXE,longitudinal}}^{\text{sphere}} = -220$ eV m µC$^{-1}$, $V_{\text{FXE,transverse}}^{\text{sphere}} = 370$ eV m µC$^{-1}$, and $V_{\text{FXE,shear}}^{\text{sphere}} = -660$ eV m µC$^{-1}$, respectively. Like the potentials, these can be simply added after multiplying by the corresponding flexoelectric coefficient.

### 3.4 Effect of Strain Gradient Elasticity

A major feature of the calculations of $\Phi_{\text{EM}}$ and $\Phi_{\text{FXE}}$ above is the rapidly varying potential from very negative to very positive at $\mathbf{r} \approx (a, z)$, where $z \lesssim 0.1a$. This is present for all indenters except the 2-D roller. In fact, this is due to the assumed pressure distributions used in the Hertz solutions. Because the pressure distributions are not smooth at the edge of contact, strain gradients in that region are artificially large. In reality, the gradients in this region will be large, but not arbitrarily so as the Hertz solutions imply. At the nanoscale, these large gradients become important. There are two main theoretical approaches to address this: couple-stress theories (Hadjesfandiari and Dargush, 2011; Yang et al., 2002) and strain gradient theories (Lam et al., 2003; Mindlin and

Eshel, 1968). The purpose of this section is not to address theoretical details that may cause one formulation to be preferable, but to show that a form of gradient elasticity is necessary for some triboelectric contact problems. Because the theory of flexoelectricity is formulated in terms of strain gradients, and relatively simple strain gradient elasticity solutions for rigid indenters of various shapes into elastic half-spaces are available (Gao and Zhou, 2013), we will proceed using solutions from elasticity theories based on strain gradients. These solutions (and gradient elasticity problems in general) require a material parameter that gives a length scale $\ell$ for the gradient term. Here, we use the value for $SrTiO_3$, which has been estimated via *ab initio* phonon calculations (Stengel, 2016). As the scale of the mechanics problem shrinks and becomes comparable to $\ell$, the influence of strain gradient mechanics increases relative to classical elasticity.

Fig. 4 shows $\Phi_{EM}$ calculations for sphere indenters using the strain gradient elasticity solutions, which are compared to the standard Hertz solutions ($\ell = 0$). As the gradient length scale parameter $\ell$ increases, the positive band bending that is seen inside the contact region in the standard solutions becomes increasingly dominate. Additionally, the extreme potential gradients are reduced, and the shape of the potentials is generally smoother. Considering these potentials act over length scales of nanometers, these solutions that include higher order elasticity terms seem to produce more physically reasonable results without abrupt potential gradients. Similar observations follow for cylinder and cone indenters (Figs. 5 and 6), though the gradient terms are perhaps even more important for these shapes that have sharp corners.

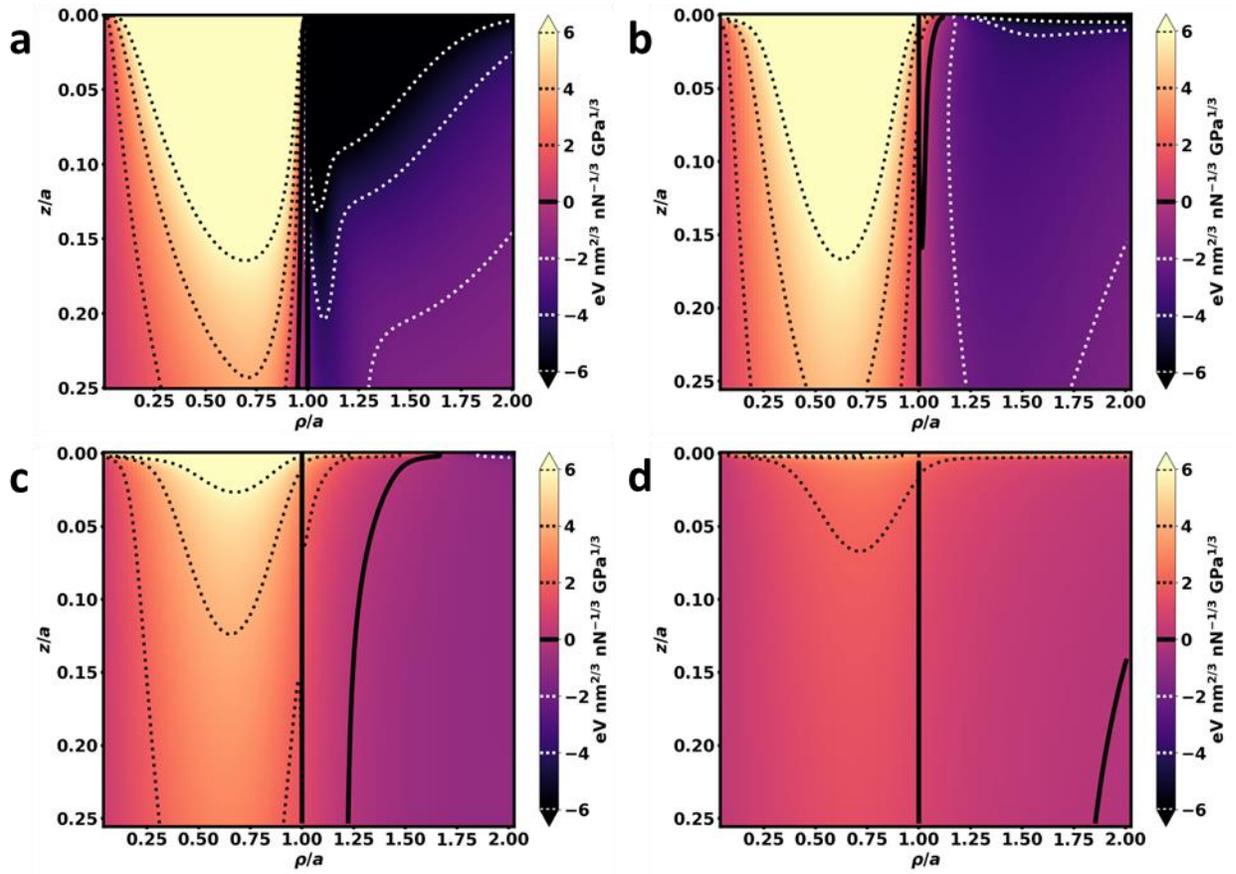

Fig. 4. Plots of $\Phi_{EM}$ for a sphere indenter. (a) Standard Hertz solution, (b-d) strain gradient elasticity solutions with (b) $\ell = a/4$, (c) $\ell = a/2$, and (d) $\ell = a$.

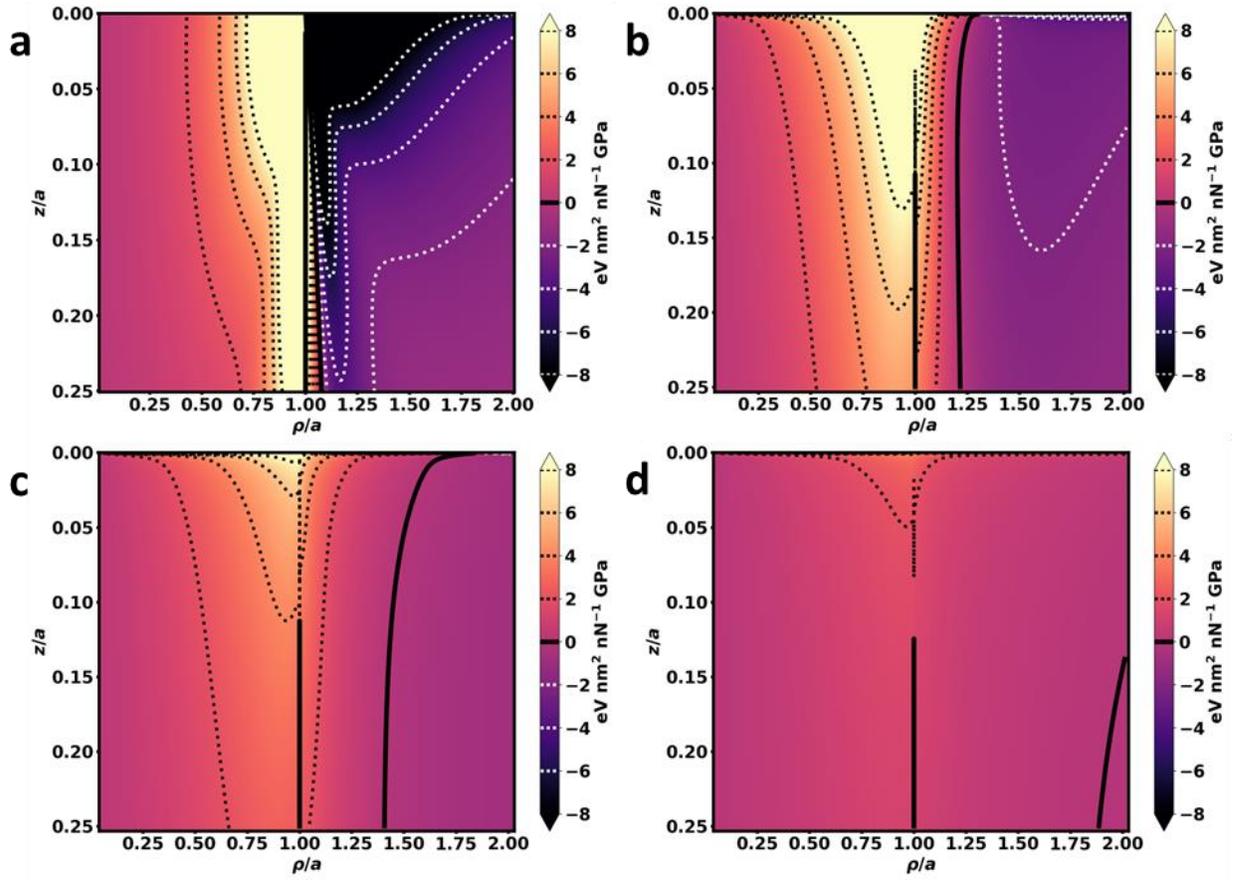

Fig. 5. Plots of $\Phi_{EM}$ for a cylinder indenter. (a) Standard Hertz solution, (b-d) strain gradient elasticity solutions with(b) $\ell = a/4$, (c) $\ell = a/2$, and (d) $\ell = a$.

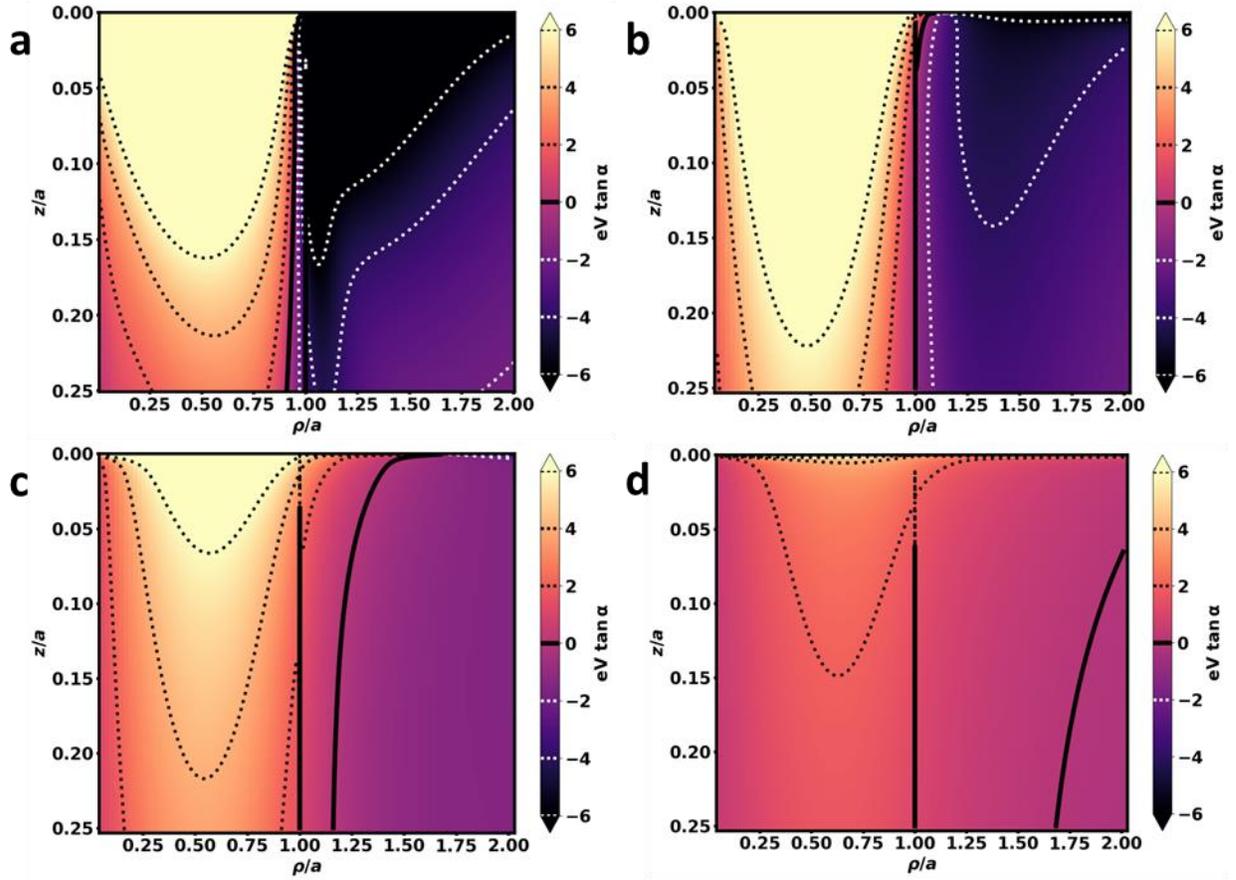

Fig. 6. Plots of $\Phi_{EM}$ for a cone indenter. (a) Standard Hertz solution, (b-d) strain gradient elasticity solutions with (b) $\ell = a/4$, (c) $\ell = a/2$, and (d) $\ell = a$.

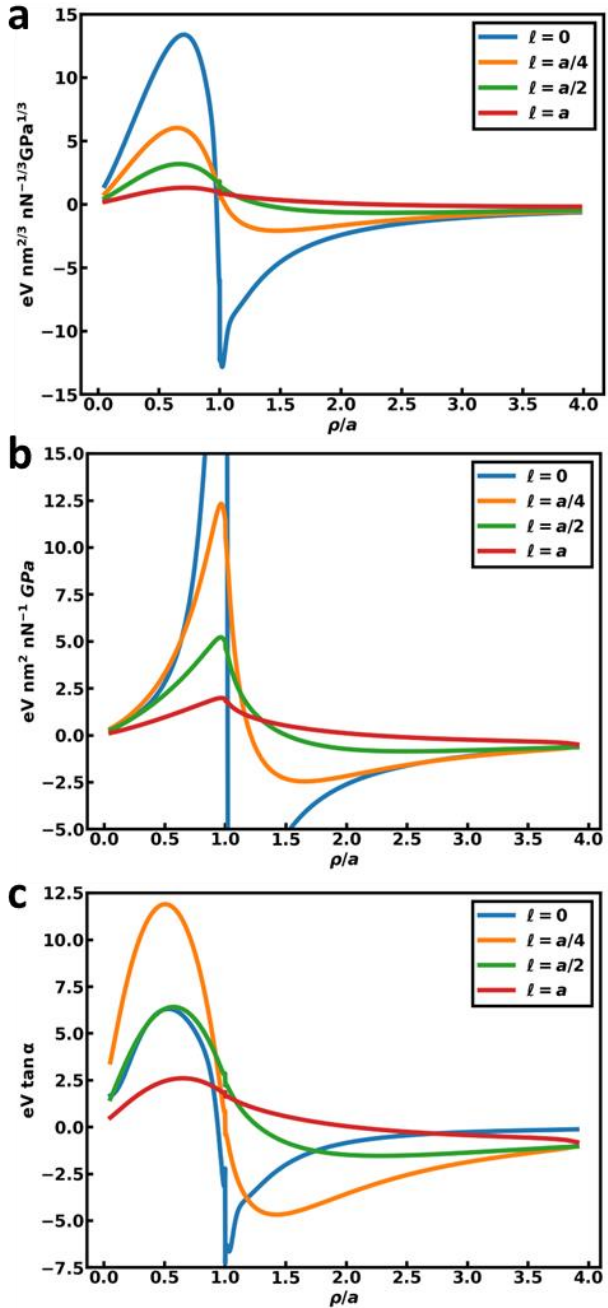

Fig. 7. Plots of $\Phi_{EM}$ sliced at $z = 0.05a$ for different values of the gradient elasticity scale $\ell$. The smoothing effect of larger $\ell$ values is evident for a (a) sphere, (b) cylinder, and (c) cone indenter.

Fig. 7 clarifies the smoothing effect of the gradient elasticity term, as a slice of $\Phi_{EM}$ at $z = 0.05a$ is plotted for different values of $\ell$. This clearly shows the smoothing is the strongest near $\rho = a$ where the Hertzian pressure distributions are not smooth.

For a particular material, e.g., $SrTiO_3$, $\ell$ is a constant, not a multiple of $a$. Therefore, the shape of the potential will change with the length scale of the indentation, i.e., $a$. For the $Pt_{0.8}Ir_{0.2}$-$SrTiO_3$ indenter system, there is little difference in the shape over a reasonable range of contact radii, as can be seen in Fig. 8. The differences are also relatively small over similar ranges for other indenter shapes (Figs. 9 and 10).

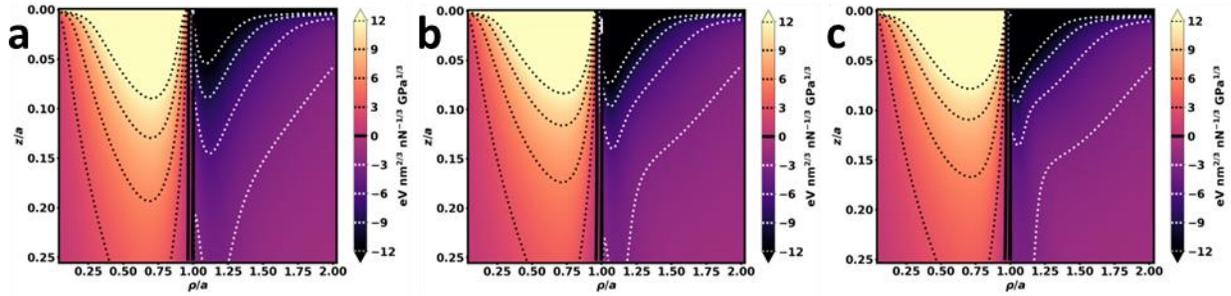

Fig. 8. Plots of $\Phi_{EM}$ for a sphere indenter with $\ell = \ell_{STO} = 4.04$ Å and contact radii (a) $a$, (b) $2a$, and (c) $4a$, where $a = 5.47$ nm is the contact radius when $F = 1$ μN and $R = 30$ nm.

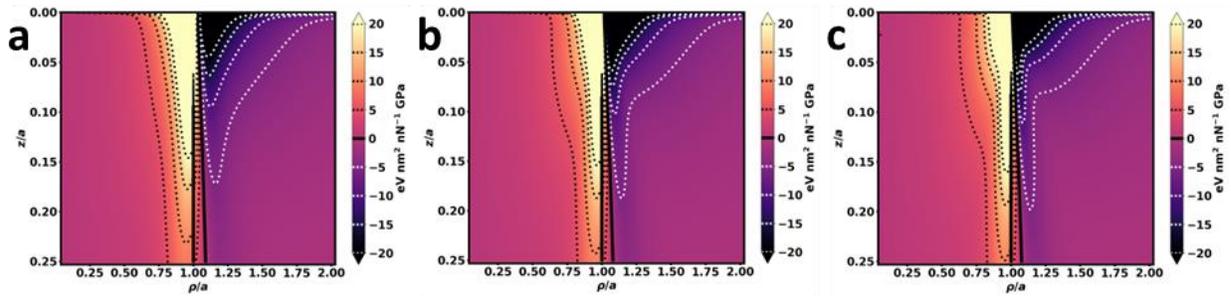

Fig. 9. Plots of $\Phi_{EM}$ for a cylinder indenter with $\ell = \ell_{STO} = 4.04$ Å and contact radii (a) $a$, (b) $2a$, and (c) $4a$, where $a = 10$ nm.

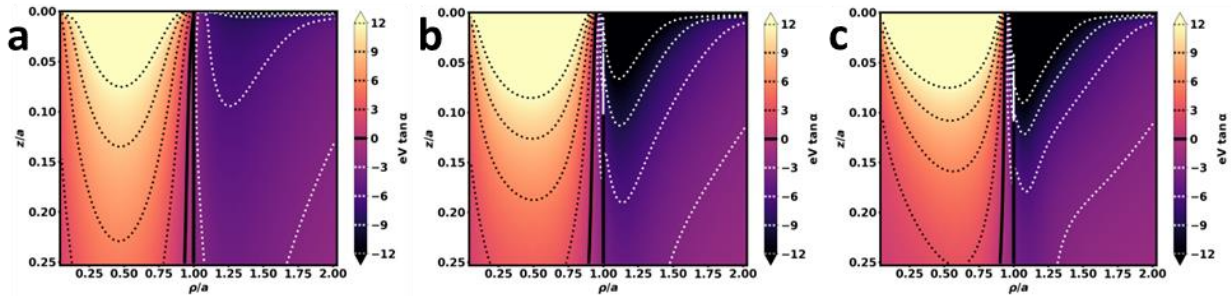

Fig. 10. Plots of $\Phi_{EM}$ for a cone indenter with $\ell = \ell_{STO} = 4.04$ Å and contact radii (a) $a$, (b) $2a$, and (c) $4a$, where $a = 2.12$ nm is the contact radius when $F = 2$ μN and $\alpha = \pi/4$.

## 4 Discussion

We have calculated normalized electromechanical potentials $\Phi_{EM}$ for a variety of rigid indenter shapes contacting an elastic half-space and described how the potentials scale with applied force and geometric parameters. These potentials are critical in understanding the electromechanical details of contact problems. Additionally, we describe how the total elastic energy and the total flexoelectric potential scale for each asperity shape.

For the 3-D cases indenting SrTiO$_3$, the region outside the contact radius, where $\rho > a$, does not depend strongly on the shape, while the inside region, where $\rho < a$, is strongly dependent on the shape. Considering triboelectric charge transfer during indentation, we note that these potentials will affect electron transport during contact. Therefore, we would expect the shape of the asperities to be a critical parameter if the region inside the contact area is important. On the other hand, if the important region is just outside the contact area, then the shape of the asperity should not be very important, with two exceptions. First, the shape changes how $\Phi_{EM}$ scales with $F$, $R$ or $\alpha$, and $Y$. Second, since the scaling of the contact radius $a$ is dependent on the shape, the indenter shape may also affect whether calculations must include strain gradient elasticity effects to be accurate over the desired range of force and length scales.

Fig. 3 shows how individual components of the flexoelectric tensor contribute to the shape of the potential. In many regions under the contact, two components have opposite effects. For example, for a larger transverse flexoelectric coefficient would induce a larger positive potential at $(\rho, z) = (0.5a, 0.05a)$, while a larger shear flexoelectric coefficient would induce a larger negative potential at the same location. This suggests that, for a particular case of interest, determining the relative magnitudes of the flexoelectric coefficients is necessary to determine the shape and even the sign of $\Phi_{FXE}$ in a region of interest.

While $\Phi_{EM}$ scales nicely with just a few parameters, for practical problems, there can be other potentials that do not scale in the same way. For example, when Nb:SrTiO$_3$ is indented by Pt$_{0.8}$Ir$_{0.2}$, the metal-semiconductor interface creates a depletion potential in the semiconductor which does not scale with the contact pressure (Olson et al., 2022). While a single conveniently-normalized solution is no longer possible, the computationally expensive calculation of $\Phi_{EM}$ is greatly simplified by using the scaling factors presented here – $\Phi_{EM}$ only needs to be calculated once for a given asperity shape before it can be used for a range of forces and sizes.

Some calculated values, especially the flexoelectric potentials given in equation 5 which can vary by as much as $\pm 20\%$, are sensitive to calculation parameters such as the resolution used in discrete integration. This can be attributed to the large, artificial discontinuity that occurs at the contact edge $\rho = a$. Since there are extreme gradients at the contact edge in the Hertz solution, the contribution of this region to the flexoelectric potential is very large and therefore sensitive to the choice of discrete points. Care is needed.

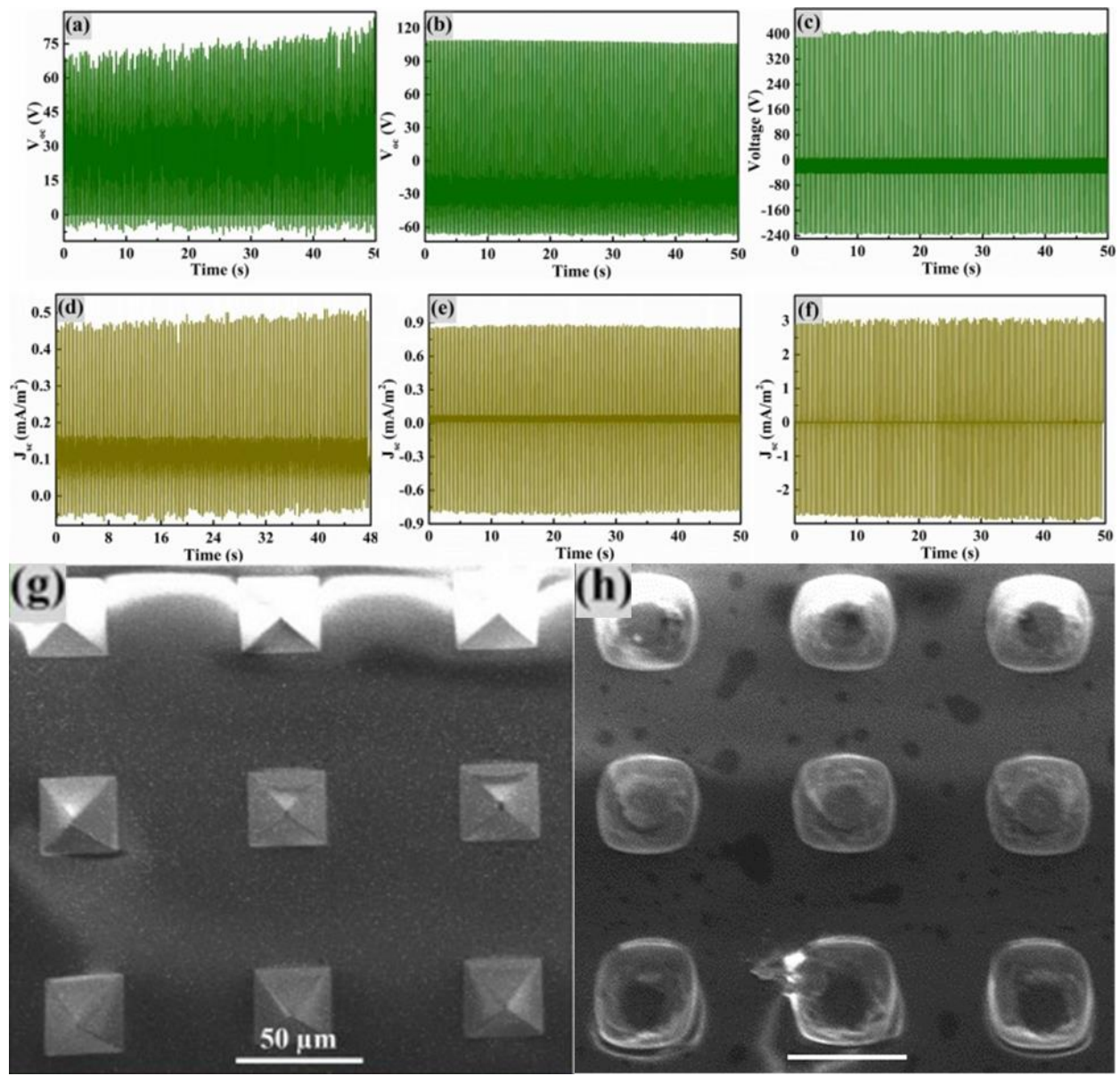

Fig. 11. (a-c, d-f) Open-circuit voltage and short-circuit current, respectively, measured for a cellulose acetate nanofiber/micro-patterned PDMS TENG, with (a,d) flat, (b,e) dome, and (c,f) pyramid micropatterns. (g,h) SEM images of (g) pyramid and (h) dome micropatterns. The scalebar in both images is 50 μm. Reprinted from Nano Energy, 98, Varghese, H., H. M. A. Hakkeem, K. Chauhan, E. Thouti, S. Pillai and A. Chandran, A high-performance flexible triboelectric nanogenerator based on cellulose acetate nanofibers and micropatterned PDMS films as mechanical energy harvester and self-powered vibrational sensor, 107339, Copyright (2022), with permission from Elsevier.

Does any of this connect to experiments? As mentioned in the introduction, experiments have been performed already with different asperity shapes and/or different roughness. The common explanation for these is that a change in the contact area is most important. However, there are problems with this. Considering the case of surface modified polydimethylsiloxane (PDMS) films

(Varghese et al., 2022), we estimate the increase in contact area using the reported measurements derived from the images reproduced in Fig. 11 (g-h). We assume that the domes are half-spheres and that the height of the pyramids is equal to their side length. Then, there is approximately a 60% and 17% increase in surface area for domes and pyramids, respectively. However, they observe an increase in short-circuit current density and open-circuit voltage of about 80% and 40% for domes, and 500% and 530% for pyramids, respectively. This disparity in magnitude and the reverse trend between domes and pyramids suggests that an explanation beyond contact area is necessary.

If we consider these results in the context of the analysis herein, this is simply explained by the difference in asperity shape. For comparison to our analysis, consider a cone with half angle $\pi/4$ (45°) and a sphere with the radius given by Varghese, 21 μm. Then, if the 3 N force is spread evenly over the artificial asperities, which have number density ~150 mm$^{-2}$ across the 2 cm$^2$ sample, and we take the modulus of the Sylgard 184 PDMS to be 1.32 MPa (Moučka et al., 2021), we can calculate the total flexoelectric potential $V_{\text{FXE}}$ for the two asperity shapes. For the pyramids, treated as cones, we have $-13.4 \text{ V} \tan\frac{\pi}{4} = -13.4 \text{ V}$. For the domes, we have $-5.5 \text{ V} \left(\frac{F}{R^2 Y}\right)^{\frac{1}{3}} = -5.5 \text{ V} \left(\frac{3 \text{ N}/(150 \text{ mm}^{-2} \cdot 2 \text{ cm}^2)}{(21 \text{ μm})^2 \, 1.32 \text{ MPa}}\right)^{\frac{1}{3}} = -3.1 \text{ V}$. This agrees very well with the relative difference between the open-circuit voltages, where that for domes is about 3.8 times smaller than for pyramids. This example of qualitative agreement with experimental data provides strong evidence for our analysis above. Simple contact area arguments will not be sufficient to explain triboelectric charge transfer – the details of the contact, including the asperity shape and size, matter.

Other experimental results also support the point that asperity shape matters. One study of Patterned-polymethylmethacrylate (PMMA)/Patterned-PDMS TENGs found that charge transfer increased in the order of flat surfaces, line-patterns (like repeating copies of this work's punch indenter), pillars (i.e., cylinders), and finally hexagonal cones (Mahmud et al., 2016). Furthermore, different sizes of pillars were measured. Smaller pillars produced larger TENG power density with approximately and inverse-square relationship to the pillar radius, in agreement with equation 3(b). Like other cases, the change in charge transfer is too large to be attributed to contact area changes. Another study has shown differences between dome and pillar morphologies in Au/Patterned-PDMS (Tcho et al., 2017).

We argue that these experimental results qualitatively validate the analysis herein.

## 5 Conclusion

The results presented here support two important points about nanoscale electromechanical contact problems. First, the shape of asperities is important to triboelectricity because it determines how the electromechanical response will scale with force, size, and modulus. The shape of the potentials that result from different asperity shapes may additionally provide insight into the mechanisms driving triboelectric charge transfer. For SrTiO$_3$ slabs, the shape of the indenter is likely important if charge transfer is driven by the potential that forms inside the contact radius. On the other hand, if the charge transfer is dependent on the potential outside the contact, the shape of the indenter is likely unimportant. For other material systems, e.g., for a material with opposite flexoelectric coefficient signs as SrTiO$_3$, the opposite may be true, or the asperity shape may be important regardless of the relevant region. Second, contact solutions that consider strain gradient elasticity seem to give more realistic $\Phi_{EM}$ by softening extreme gradients that result from standard Hertzian contact solutions. This becomes increasingly significant at contacts of smaller length scales, as well as for materials where strain gradient elasticity is more important.

The variations with asperity shape are qualitatively consistent with experimental evidence.

# References

Aazem, I., Walden, R., Babu, A., Pillai, S.C., 2022. Surface patterning strategies for performance enhancement in triboelectric nanogenerators. Results Eng. 16, 100756.

Alamos, F.J., Philo, M., Go, D.B., Schmid, S.R., 2021. Asperity contact under creep conditions. Tribol. Int. 160, 107039.

Bhushan, B., 1998. Contact mechanics of rough surfaces in tribology: multiple asperity contact. Tribol. Lett. 4, 1-35.

De Pellegrin, D.V., Stachowiak, G.W., 2004. Evaluating the role of particle distribution and shape in two-body abrasion by statistical simulation. Tribol. Int. 37, 255-270.

Fan, F.-R., Tian, Z.-Q., Lin Wang, Z., 2012. Flexible triboelectric generator. Nano Energy 1, 328-334.

Fischer-Cripps, A.C., 2007. Elastic Indentation Stress Fields, in: Ling, F.F. (Ed.), Introduction to Contact Mechanics, 2 ed. Springer US, New York, pp. 77-100.

Gao, X.-L., Zhou, S.-S., 2013. Strain gradient solutions of half-space and half-plane contact problems. Z. Angew. Math. Phys. 64, 1363-1386.

Greenwood, J.A., Williamson, J.B.P., Bowden, F.P., 1966. Contact of nominally flat surfaces. Proc. R. Soc. Lond. A. Math. Phys. Sci. 295, 300-319.

Hadjesfandiari, A.R., Dargush, G.F., 2011. Couple stress theory for solids. Int. J. Solids Struct. 48, 2496-2510.

Hertz, H., 1882. Ueber die Berührung fester elastischer Körper. J. Reine Angew. Math. 1882, 156-171.

Hong, J.W., Vanderbilt, D., 2013. First-principles theory and calculation of flexoelectricity. Phys. Rev. B 88, 174107.

Huang, J., Fu, X., Liu, G., Xu, S., Li, X., Zhang, C., Jiang, L., 2019. Micro/nano-structures-enhanced triboelectric nanogenerators by femtosecond laser direct writing. Nano Energy 62, 638-644.

Johnson, K.L., 1982. One Hundred Years of Hertz Contact. Proc. Inst. Mech. Eng. 196, 363-378.

Johnson, K.L., 1985. Contact Mechanics. Cambridge University Press, Cambridge.

Kim, W.-G., Kim, D.-W., Tcho, I.-W., Kim, J.-K., Kim, M.-S., Choi, Y.-K., 2021. Triboelectric Nanogenerator: Structure, Mechanism, and Applications. Acs Nano 15, 258-287.

Lacks, D.J., Shinbrot, T., 2019. Long-standing and unresolved issues in triboelectric charging. Nat. Rev. Chem. 3, 465-476.

Lam, D.C.C., Yang, F., Chong, A.C.M., Wang, J., Tong, P., 2003. Experiments and theory in strain gradient elasticity. J. Mech. Phys. Solids 51, 1477-1508.

Liang, S.-C., Zhang, J.-P., Fan, L.-S., 1996. Electrostatic Characteristics of Hydrated Lime Powder during Transport. Ind. Eng. Chem. Res. 35, 2748-2755.
M'Ewen, E., 1949. XLI. Stresses in elastic cylinders in contact along a generatrix (including the effect of tangential friction). Philos. Mag. 40, 454-459.
Mahmud, M.A.P., Lee, J., Kim, G., Lim, H., Choi, K.-B., 2016. Improving the surface charge density of a contact-separation-based triboelectric nanogenerator by modifying the surface morphology. Microelectron. Eng. 159, 102-107.
Mills, C.M., Ester, V.C., Henkin, H., 1956. Measurement of Static Charge on Hair. J. Soc. Cosmet. Chem. 7, 466-475.
Mindlin, R.D., Eshel, N.N., 1968. On first strain-gradient theories in linear elasticity. Int. J. Solids Struct. 4, 109-124.
Mishra, S.K., Sana, T., 2022. Mitigating massive triboelectric charging of drill in shadowed region of Moon. Mon. Not. R. Astron. Soc. 512, 4730-4735.
Mizzi, C.A., Lin, A.Y.W., Marks, L.D., 2019. Does Flexoelectricity Drive Triboelectricity? Phys. Rev. Lett. 123, 116103.
Mizzi, C.A., Marks, L.D., 2021. The role of surfaces in flexoelectricity. J. Appl. Phys. 129, 224102.
Mizzi, C.A., Marks, L.D., 2022. When Flexoelectricity Drives Triboelectricity. Nano Letters 22, 3939-3945.
Moučka, R., Sedlačík, M., Osička, J., Pata, V., 2021. Mechanical properties of bulk Sylgard 184 and its extension with silicone oil. Scientific Reports 11, 19090.
Muthu, M., Pandey, R., Wang, X., Chandrasekhar, A., Palani, I.A., Singh, V., 2020. Enhancement of triboelectric nanogenerator output performance by laser 3D-Surface pattern method for energy harvesting application. Nano Energy 78, 105205.
Naik, S., Hancock, B., Abramov, Y., Yu, W., Rowland, M., Huang, Z., Chaudhuri, B., 2016. Quantification of Tribocharging of Pharmaceutical Powders in V-Blenders: Experiments, Multiscale Modeling, and Simulations. J. Pharm. Sci. 105, 1467-1477.
Olson, K.P., Mizzi, C.A., Marks, L.D., 2022. Band Bending and Ratcheting Explain Triboelectricity in a Flexoelectric Contact Diode. Nano Letters 22, 3914-3921.
Pan, S.H., Zhang, Z.N., 2019. Fundamental theories and basic principles of triboelectric effect: A review. Friction 7, 2-17.
Popov, V.L., Heß, M., Willert, E., 2019. Normal Contact Without Adhesion, in: Popov, V.L., Heß, M., Willert, E. (Eds.), Handbook of Contact Mechanics: Exact Solutions of Axisymmetric Contact Problems. Springer Berlin Heidelberg, Berlin, Heidelberg, pp. 5-66.
Siripuram, R.B., Stephens, L.S., 2004. Effect of Deterministic Asperity Geometry on Hydrodynamic Lubrication. J. Tribol. 126, 527-534.
Steinpilz, T., Joeris, K., Jungmann, F., Wolf, D., Brendel, L., Teiser, J., Shinbrot, T., Wurm, G., 2019. Electrical charging overcomes the bouncing barrier in planet formation. Nat. Phys. 16, 225-229.
Stengel, M., 2015. From flexoelectricity to absolute deformation potentials: The case of SrTiO3. Phys. Rev. B 92, 205115.
Stengel, M., 2016. Unified ab initio formulation of flexoelectricity and strain-gradient elasticity. Phys. Rev. B 93, 245107.
Tcho, I.W., Kim, W.G., Jeon, S.B., Park, S.J., Lee, B.J., Bae, H.K., Kim, D., Choi, Y.K., 2017. Surface structural analysis of a friction layer for a triboelectric nanogenerator. Nano Energy 42, 34-42.

Varghese, H., Hakkeem, H.M.A., Chauhan, K., Thouti, E., Pillai, S., Chandran, A., 2022. A high-performance flexible triboelectric nanogenerator based on cellulose acetate nanofibers and micropatterned PDMS films as mechanical energy harvester and self-powered vibrational sensor. Nano Energy 98, 107339.
Watanabe, H., Ghadiri, M., Matsuyama, T., Ding, Y.L., Pitt, K.G., Maruyama, H., Matsusaka, S., Masuda, H., 2007. Triboelectrification of pharmaceutical powders by particle impact. Int. J. Pharm. 334, 149-155.
Williams, M.W., 2012. Triboelectric charging of insulating polymers–some new perspectives. AIP Adv. 2, 010701.
Wong, J., Kwok, P.C.L., Chan, H.-K., 2015. Electrostatics in pharmaceutical solids. Chem. Eng. Sci. 125, 225-237.
Yamamoto, T., Suzuki, S., Kawaguchi, K., Takahashi, K., 1998. Temperature dependence of the ideality factor of Ba1-xKxBiO3/Nb-doped SrTiO3 all-oxide-type Schottky junctions. Jpn. J. Appl. Phys. 37, 4737-4746.
Yang, F., Chong, A.C.M., Lam, D.C.C., Tong, P., 2002. Couple stress based strain gradient theory for elasticity. Int. J. Solids Struct. 39, 2731-2743.
Zhang, X.-W., Li, G.-Z., Wang, G.-G., Tian, J.-L., Liu, Y.-L., Ye, D.-M., Liu, Z., Zhang, H.-Y., Han, J.-C., 2018. High-Performance Triboelectric Nanogenerator with Double-Surface Shape-Complementary Microstructures Prepared by Using Simple Sandpaper Templates. ACS Sustain. Chem. Eng. 6, 2283-2291.
Zou, Y., Xu, J., Chen, K., Chen, J., 2021. Advances in Nanostructures for High-Performance Triboelectric Nanogenerators. Adv. Mater. Technol. 6, 2000916.

**Funding**

This work was supported by Northwestern University McCormick School of Engineering.